\documentclass[twocolumn]{aastex62}

\usepackage{epsf}
\usepackage{amsmath}                % American Mathematical Society package
\usepackage{amsfonts}               % American Mathematical Society fonts
\usepackage{amssymb}                % American Mathematical Society symbol
\usepackage{epsfig}                 % EPS figures
\usepackage{float}
\usepackage{color}
\usepackage{tabularx}
\usepackage{comment}
\usepackage{makecell}
\usepackage{enumitem}               %To enumarate cases
\def \s{~\rm{s}}
\def \km{~\rm{km}}

\def \erg{~\rm{erg}}

\begin{document}

\title{Common envelope jets supernova with thermonuclear outburst progenitor for the enigmatic supernova remnant W49B}

\author{Aldana Grichener}
\affiliation{Department of Physics, Technion, Haifa, 3200003, Israel; aldanag@campus.technion.ac.il; soker@physics.technion.ac.il}

\author[0000-0003-0375-8987]{Noam Soker}

\affiliation{Department of Physics, Technion, Haifa, 3200003, Israel; aldanag@campus.technion.ac.il; soker@physics.technion.ac.il}

\begin{abstract}
We suggest a common envelope jets supernova (CEJSN) origin to the supernova remnant (SNR) W49B where jets launched by a neutron star (NS) that collapsed to a black hole (BH) together with a thermonuclear outburst of the disrupted red super giant's (RGS's) core powered and shaped the ejecta. The jets account for the highly non-spherical morphology of W49B and the thermonuclear outburst to its high iron abundance. CEJSNe are violent events powered by jets that a NS or a BH launch as they orbit inside a red supergiant star and accrete mass from its envelope and then from its core. We classify the CEJSN process to either a case where the NS/BH enters the core to form a common envelope evolution (CEE) inside the core or to a case where the NS/BH tidally disrupts the core. In the later case the core material forms an accretion disk around the NS that might experience a thermonuclear outburst, leading to an energetic event powered by both jets and thermonuclear burning. We term this scenario thermonuclear CEJSN. 
We find that the maximum core mass that leads to this scenario with a NS is $2 M_{\rm \odot} \lesssim M_{\rm core} \lesssim 3.5M_{\rm \odot}$. We estimate the event rates of CEJSN that go through tidal disruption of the core by a NS to be 5 per 1000 core collapse supernovae.
\end{abstract}


% ==========================================================
\section{INTRODUCTION}
\label{sec:intro}
% ==========================================================

A \textit{common envelope jet supernova (CEJSN)} is a gravitational-powered transient event with a kinetic energy that might reassemble or exceed the energy of a typical supernova (SN) explosion. In a CEJSN event a neutron star (NS) or a black hole (BH) spirals-in inside a red supergiant (RSG) star and accretes a fraction of the mass in its envelope and in its core through an accretion disk that launches part of the matter as two opposite jets (e.g., \citealt{ArmitageLivio2000, Chevalier2012, Papishetal2015, SokerGilkis2018,  GrichenerSoker2019a, LopezCamaraetal2019, LopezCamaraetal2020MN, Schreieretal2021, Soker2021,GrichenerCohenSoker2021, Hilleletal2023turb}). The collision of the jets with the RSG star and possibly with the circumstellar material (CSM) expels large amounts of mass and channels a fraction of the kinetic energy to radiation. Cases where the NS/BH does not mange to reach the core of the RSG star are termed \textit{CEJSN impostors} \citep{Gilkisetal2019}, and are characterized by lower energies. Jets are a key ingredient in CEJSNe/CEJSN impostors. They distinguish them from closely related scenarios of common envelope evolution (CEE) with NS/BH companions that assume other energy transfer mechanisms (e.g., \citealt{ThorneZytkow1977, FryerWoosley1998, ZhangFryer2001, BarkovKomissarov2011, Thoneetal2011, Schroderetal2020}). 

CEJSN and CEJSN-impostor events might account for observations that are classified as peculiar core collapse supernovae (CCSNe). \cite{SokerGilkis2018} attribute the enigmatic SN iPTF14hls to a CEJSN event; \cite{Dongetal2021} explain the luminous radio transient VTJ121001+495647 by a CEJSN scenario; \cite{SokeretalGG2019} propose the polar CEJSN channel for the AT2018cow transient while \cite{Soker2022FBOT} and \cite{CohenSoker2023} consider a CEJSN-impostor scenario for said event. Moreover, \cite{Schroderetal2020} suggest that the unusual transients SN1979c and SN1998s might result from explosions driven from the merger of a NS/BH with the core of a giant star.

The different phases during the RSG-NS/BH CEE in which the jets are launched can lead to different outcomes. CEJSNe where the NS accretes mass while spiraling-in inside the CO core of an RSG might be one of the sites for r-process nucleosynthesis \citep{GrichenerSoker2019a, GrichenerSoker2019b, Gricheneretal2022, GrichenerSoker2022, Grichener2023}. Jets that a BH launches as it orbits deep in the envelope of an RSG star (\citealt{GrichenerSoker2021}; \citealt{Grichener2023}) or just before it enters the CEE \citep{SridharMetzger2022, Sridharetal2023} might be one of the sources of the PeV neutrinos detected by IceCube (e.g., \citealt{Aartsenetal2013}).   

In the present study we consider CEJSN events where a NS companion tidally disrupts the core of an RSG to form a massive accretion disk around the NS that experiences a thermonuclear outburst. In section \ref{sec:TDE} we derive the condition for this channel, rather than a channel where the NS continues to a CEE inside the core, and estimate its event rate. In section \ref{sec:W49B} we suggest that the enigmatic supernova remnant (SNR) W49B results from a CEJSN with a thermonuclear outburst. We summarize our results and discuss their implications in section \ref{sec:Summary}.

% ==========================================================
\section{Tidal disruption of the core}
\label{sec:TDE}
% ==========================================================

% ==========================================================
\subsection{Criteria for tidal disruption of the core}
\label{subsec:TDEcriteria}
% ==========================================================

Prior to the CEJSN event, the NS/BH spirals-in inside the envelope of the RSG. The spiraling-in  lasts for several months to few years and powers a long transient event by the release of orbital energy and by the collision of the jets with the envelope's matter. The CEE might end in one of three ways. 
\begin{enumerate}
\item NS/BH-core CEE (e.g., \citealt{GrichenerSoker2019a}). The NS/BH enters the core of the RSG and spirals-in inside it, launching jets that eventually destroy the core. The jets that collide with the core and the earlier ejecta power a days to weeks-long superluminous event.  
\item Tidal disruption of the core (subject of this paper). The NS/BH tidally disrupts the core before it manages to reach the core's surface. The material of the destroyed core is ignited by thermonuclear burning and expelled through jets, leading to a bright transient. We note that this scenario is possible even if the NS/BH does not launch jets during the earlier CEE as some studies suggest (e.g., \citealt{Murguia-Berthieretal2017}) or if the jets are weak due to jet feedback mechanism (e.g., \citealt{Soker2016Rev, GrichenerCohenSoker2021,  Hilleletal2022FB, LopezCamaraetal2022}).
\item CEJSN impostor (e.g., \citealt{Gilkisetal2019}). The NS/BH expels the entire envelope before it reaches the core, leaving it intact. This transient does not reach the high luminosity of the final CEJSN phase where the NS/BH is inside the core. The surviving core might eventually explode and leave behind a NS or a BH, forming a double compact object binary system. If the binary remains bound it might later merge through emission of gravitational waves (e.g., \citealt{VignaGomezetal2018}; \citealt{Belczynskietal2020}; \citealt{Zevinetal2020}; \citealt{Grichener2023}).  
\end{enumerate}

%In even rarer cases such a scenario might take place with a BH companion.
Here we focus on CEJSNe with a NS companion since we suggest the SNR W49B originates from a tidal disruption of the core by a NS in a CEJSN event (section \ref{sec:W49B}). The immense gravity of the NS might tidally destroy the core (outcome 2) at a radius of 
\begin{equation}
%\begin{split}
r_{\rm t}= f R_{\rm core} \left(\frac{M_{\rm NS}}{M_{\rm core}}\right)^{1/3},
%\\
%\end{split}
\label{eq:TidalRadius}
\end{equation}
where $R_{\rm core}$ and $M_{\rm core}$ are the radius and the mass of the core, respectively, and $M_{\rm NS}$ is the mass of the NS. $f$ is an order of unity disruption parameter that accounts for different effects of the objects in a binary system where a tidal disruption event occurs (e.g., \citealt{Harris1996}; \citealt{Fryeretal1998}; \citealt{GuillochonRamirezRuiz2013}; \citealt{Gaftonetal2015}; \citealt{LawSmithetal2017}; \citealt{GaftonRosswog2019}; \citealt{Nixonetal2021}). The criteria for a NS to tidally destroy the core and form a disk around it rather than to plunge inside it initiating a CEE phase (in which the jets will eventually destroy the core) is $r_{\rm t} \gtrsim R_{\rm core}$. Subsisting this in equation (\ref{eq:TidalRadius}) yields 
\begin{equation}
%\begin{split}
M_{\rm core} \lesssim  f^{3}M_{\rm NS} = 2.8 M_{\rm \odot} \left(\frac{f^{3}}{2}\right) \left(\frac{M_{\rm NS}}{1.4 M_{\rm \odot}}\right),
%\\
%\end{split}
\label{eq:TidalConditionMass}
\end{equation}
where we calibrate the equation with a disruption parameter $f=1.26$ taken from \cite{Harris1996} assuming the NS-core systems does not have time to reach synchronization. We also consider higher and lower values of the disruption parameter around this value based on the studies above. 

We use the stellar evolution code \textsc{mesa} (Modules for
Experiments in Stellar Astrophysics; \citealt{Paxtonetal2011}; \citealt{Paxtonetal2013}; \citealt{Paxtonetal2015}; \citealt{Paxtonetal2018}; \citealt{Paxtonetal2019}) version r22.05.1 to follow the evolution of massive stars and find the critical initial RSG mass whose core satisfies the condition in equation (\ref{eq:TidalConditionMass}). We run stellar models of masses in the range $11 \lesssim \left(\frac{M_{\rm ZAMS}}{M_{\rm \odot}}\right) \lesssim 20$ with solar metalicity ($Z_{\rm \odot}=0.0142$; \citealt{Asplundetal2009}). The rest of the code parameters are the same as in the default \textit{test-suite} of $20M_{\rm \odot}$ mass ($ \rm 20M\_pre\_MS\_to\_cc$) \footnote{Runs performed with the default \textit{test-suite} of $12M_{\rm \odot}$ (($ \rm 12M\_pre\_MS\_to\_cc$)) yield the same result for the core masses.}.  We assume that the NS does not affect the core until it gets close enough to destroy it, and hence we can run uninterrupted models of the \textsc{mesa} single star module.

\begin{figure}
\begin{center}
\vspace*{-5.9cm}
\hspace*{-2.7cm}
\includegraphics[width=0.79\textwidth]{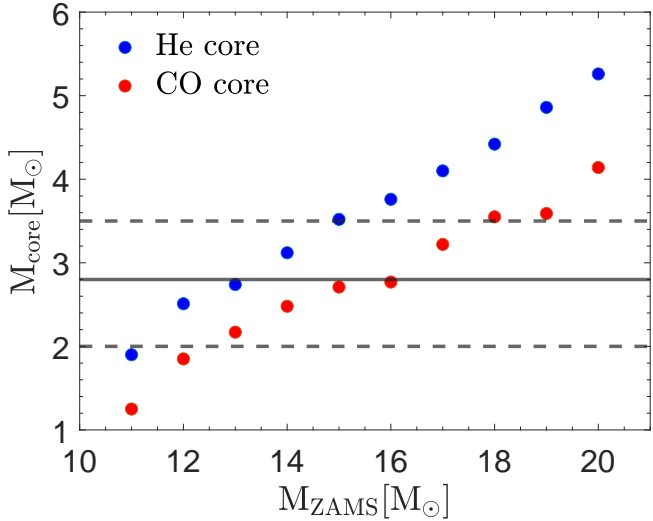}
\vspace*{-5.5cm}
\caption{He core (blue dots) and CO core (red dots) masses at the first and second expansion peaks, respectively, vs the initial mass of the RSG. The solid grey line represents the upper bound for tidal disruption of the core according to \cite{Harris1996}. The dashed grey lines represent this bound for the highest and the lowest values of $f$ we consider.}
\label{fig:MvsMcore}
 \end{center}
 \end{figure}

Massive stars usually exhibit two distinct expansion phases, after the exhaustion of hydrogen and helium (He) in their core, respectively. We define the He/carbon-oxygen (CO) core boundary as the outermost location where the mass fraction of hydrogen/He is smaller than 0.1, and find the masses of the He (blue dots in Fig. \ref{fig:MvsMcore}) and CO (red dots in Fig. \ref{fig:MvsMcore}) cores at the first and second expansion peaks, respectively, in our stellar models. The solid grey line marks the highest core mass that will result in a tidal disruption merger for $f^{3}=2$ \citep{Harris1996}. Due to the uncertainty in $f$ we also consider the limiting values of the core mass for a tidal disruption parameter in the range $1.5 \le f^{3} \leq 2.5 $. For $M_{\rm NS}=1.4 M_\odot$ this gives core masses between $2M_{\rm \odot}$ and $3.5M_{\rm \odot}$. We mark the boundaries of this range by the two horizontal dashed lines in Fig. \ref{fig:MvsMcore}. Our core masses are larger than in some other stellar evolution simulations that use different codes (e.g., \cite{SukhboldWoosleyHeger}). This could be due to different parameters, such as convective overshooting, or due to different definitions of the cores boundaries.
% FFFFFFFFFFFFFFFFFFFFFFFFFFFFFFFFFFFFFFFFFFFFFFFF
% FFFFFFFFFFFFFFFFFFFFFFFFFFFFFFFFFFFFFFFFFFFFFFFF

We find that for the lowest value of $f$ we consider NS-core mergers will occur through tidal disruption of the core for giants with initial masses of $M_{\rm ZAMS} \lesssim M_{\rm crit} \simeq 11M_{\rm \odot}$ in the case of an He core and for giants with initial masses of $M_{\rm ZAMS} \lesssim  M_{\rm crit} \simeq 12M_{\rm \odot}$ in the case of a CO core. The highest value of $f$ allows for tidal disruption of an He core for $M_{\rm ZAMS} \lesssim  M_{\rm crit} \simeq 15M_{\rm \odot}$ and tidal disruption of a CO core for $ M_{\rm ZAMS} \lesssim M_{\rm crit} \simeq 17.5M_{\rm \odot}$. We conclude that there is a non-negligible fraction of the parameter space for which CEJSN events end with the tidal disruption of the RSG core. Bases on the masses we found here, we turn to calculate the rate of such events.

% ==========================================================
\subsection{Event rate of the core's tidal disruption}
\label{subsec:EventRate}
% ==========================================================
\cite{Grichener2023} recently performed population synthesis of massive stars following the mergers of NSs and BHs with cores of giants using the population synthesis code \textsc{compas}. Here we use the data of these simulations for a common envelope efficiency parameter of $\alpha_{\rm CE}=1$ and the critical initial RSG mass derived in the previous subsection to find the event rates of NS-core mergers that result in tidal disruption of the core. We then compare our results to the merger rates inferred from simulations with other values of common envelope efficiency parameters.

Even though jets that the NS launches while spiralling-in inside the envelope of the RSG deposit part of their kinetic energy into the envelope and facilitate mass removal, we believe $\alpha_{\rm CE}=1$ is suitable for our scenario. The reason is that without jets $\alpha_{\rm CE} < 1$, mainly because convection in the envelope seems to be very efficient in transferring energy outwards were it is radiated away (e.g., \citealt{Gricheneretal2018, WilsonNordhaus2019, WilsonNordhaus2020, WilsonNordhaus2022}). Moreover, the jets that the NS launches induce further convection in the envelope, increasing the efficiency of convective energy transport (\citealt{Hilleletal2023turb}). In addition, the jets expel some fraction of the envelope mass at velocities above the escape velocity, such that the ejected envelope takes an energy larger than its binding energy (e.g., \citealt{Hilleletal2023turb}), reducing the efficiency of mass removal. 

\begin{figure*}
\begin{center}
%\vspace*{-5.9cm}
%\hspace*{-2.7cm}
\includegraphics[width=0.9\textwidth]{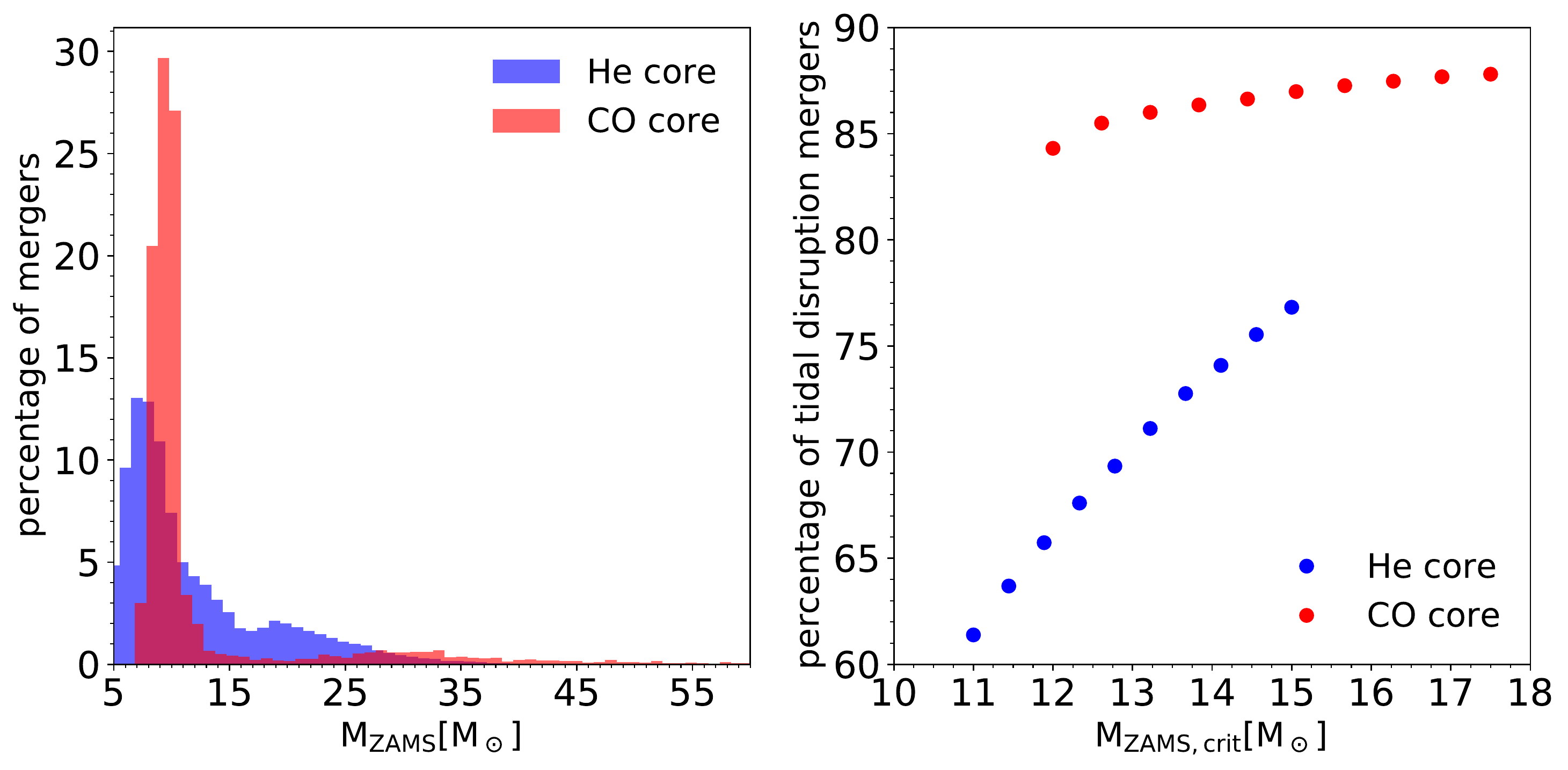}
%\vspace*{-5.5cm}
\caption{Left panel: Initial mass distribution of the giant stars in binaries that result in NS-core mergers (either via tidal disruption or CEE of a NS inside the core) for a common envelope efficiency parameter of $\alpha_{\rm CE}=1$. The blue and red bins represent giants in which the merger occurred when the core was mainly composed of He and of CO, respectively. Each distribution is normalized relative to itself. The width of the bins equals $1 M_{\rm \odot}$. Right panel: Percentage of systems that result in tidal disruption of the He (blue dots) or CO (red dots) core from all NS-He/CO core mergers.} 
\label{fig:mass_hist_percent_1.0}
 \end{center}
 \end{figure*}

In the left panel of Fig. \ref{fig:mass_hist_percent_1.0} we present the ZAMS mass distribution of RSG progenitors that lead to CEJSN events where a NS companion merges with the core, either through a tidal disruption of the core by the NS or via a CEE of the NS inside the core. The blue histogram is for cases where the NS merges with an He core while the red histogram represents NS-CO cores mergers. The sum over the columns of each histogram separately equals a hundred percent. We can see that in both cases there is a large concentration of stars in the regime of tidal disruption driven mergers. 
% FFFFFFFFFFFFFFFFFFFFFFFFFFFFFFFFFFFFFFFFFFFFFFFF
% FFFFFFFFFFFFFFFFFFFFFFFFFFFFFFFFFFFFFFFFFFFFFFFF

Using the results that we present in Fig. \ref{fig:MvsMcore} to analyse the mass distribution in the left panel of Fig. \ref{fig:mass_hist_percent_1.0}, we find the percentage of systems that go through tidal disruption of the core by a NS of mass $M_{\rm NS}=1.4 M_\odot$. As there is uncertainty in the value of $f$ that leads to an uncertainty in the initial RSG masses of said systems, we consider the range of possible values marked with the dashed grey lines in  Fig. \ref{fig:MvsMcore}. In the right panel of Fig. \ref{fig:mass_hist_percent_1.0} we present the percentage of the NS-core merger events that end with tidal disruption rather than CEE as function of the critical ZAMS mass of the RSG. The blue dots are for cases where the core is composed mostly of He while the red dots are for tidal disruption of CO cores. 

Combining the data from the right panel of Fig. \ref{fig:mass_hist_percent_1.0} with the total percentage of NS-core mergers from CCSNe (Fig. 7 in \citealt{Grichener2023}) we find that there are about 5 core tidal disruption events per 1000 CCSNe. From the percentage of mergers with He or CO cores (Table 2 in \citealt{Grichener2023}) we conclude that there are about 4 tidal disruption events of the RSG He core by the NS per 1000 CCSNe, and about one event per 1000 CCSNe in the case of a CO core. The latter leaves 0.2 CEJSN events where the NS enters a CEE inside the core per 1000 CCSNe, in which r-process nucleosynthesis can occur \citep{GrichenerSoker2019a}. Performing the same computation for other values of $\alpha_{\rm CE}$ we find that the rate does not change much up to $\alpha_{\rm CE}=1.75$. We note that a CEJSN where the core gets tidally disrupted by a BH could also happen. However, it would be much more rare than in the case of a NS due to a lower event rate of CEJSN with BH companions.

\begin{figure}
\begin{center}
%\vspace*{-5.9cm}
\hspace*{-0.7cm}
\includegraphics[width=0.53\textwidth]{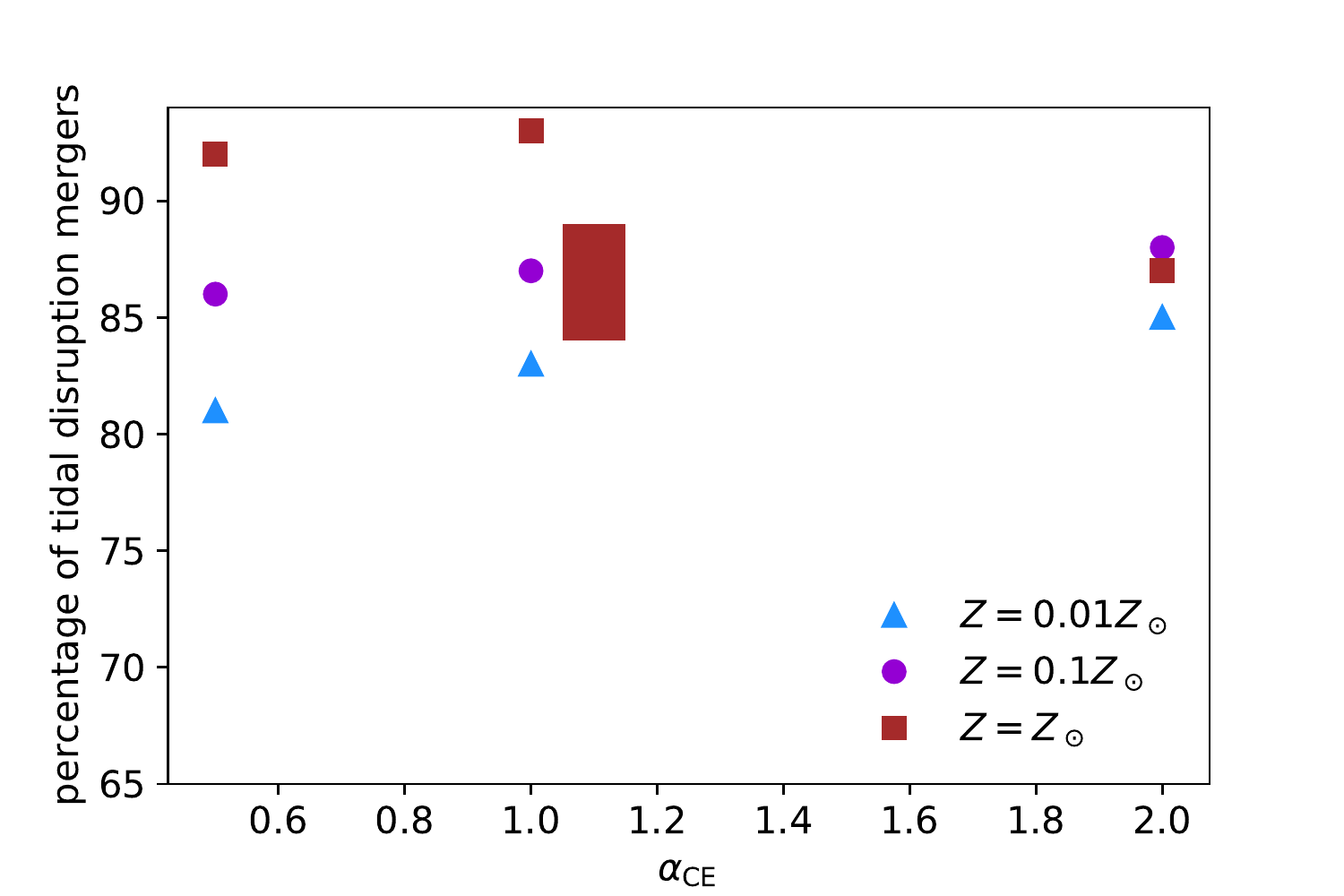}
%\vspace*{-5.5cm}
\caption{Percentage of NS-core mergers that occur through tidal disruption of the core where core masses are inferred from our \textsc{compas} simulations for different values of the efficiency parameter $\alpha_{\rm CE}$ in the case where all stars in the binary begin with metallicities of $Z=0.01Z_{\rm \odot}$ (blue dots),$Z=0.1Z_{\rm \odot}$ (purple dots) and $Z=Z_{\rm \odot}$ (brown dots) ($Z_{\rm \odot}=0.0142$; \citealt{Asplundetal2009}). The brown rectangle represents the range of results for $\alpha_{\rm CE}=1$ and $Z=Z_{\rm \odot}$ where the (CO) core masses are from \textsc{mesa}, i.e., the range of the red dotes in the right panel of  Fig. \ref{fig:mass_hist_percent_1.0}. We displaced the rectangle to the right of $\alpha_{\rm CE}=1$ for clarity.  }
\label{fig:TDEalphaZed}
 \end{center}
 \end{figure}

The stellar evolution code \textsc{mesa} solves the stellar equations in details while the population synthesis code \textsc{compas} uses approximated parametrizations. However, in this case there is an advantage to using \textsc{compas} for the connection between the mass of the core and the ZAMS mass from considerations of self-consistency. Therefore, below we use the condition of equation (\ref{eq:TidalConditionMass}) with the core mass that \textsc{compas} gives instead of the core mass from Fig. \ref{fig:MvsMcore}. In Fig. \ref{fig:TDEalphaZed} we present the percentage of NS-core mergers that occur through tidal disruption of the core when the core mass is taken from \textsc{compas}. We present our results for three different values of the CEE efficiency parameter $\alpha_{\rm CE}$ and for three different values of metallicity due to the large uncertainties in these parameters. We also mark on Fig. \ref{fig:TDEalphaZed} the distribution of the results from the CO core from Fig. \ref{fig:mass_hist_percent_1.0}, where the core mass is from \textsc{mesa} according to Fig. \ref{fig:MvsMcore}. We find that within the uncertainty of the population synthesis models and the uncertainty in the value of $\rm f$ in equation (\ref{eq:TidalConditionMass}), the results presented in the right panel of Fig. \ref{fig:mass_hist_percent_1.0} and in Fig. \ref{fig:TDEalphaZed} are in agreement when we consider CO cores.  
% FFFFFFFFFFFFFFFFFFFFFFFFFFFFFFFFFFFFFFFFFFFFFFFF
% FFFFFFFFFFFFFFFFFFFFFFFFFFFFFFFFFFFFFFFFFFFFFFFF

Similarly to most population synthesis codes, \textsc{compas} uses parametrizations to simplify complicated stellar evolution processes and stages. The results we present in Figs. \ref{fig:mass_hist_percent_1.0} and \ref{fig:TDEalphaZed} depend on the initial distributions of stellar and binary properties as well as on the way \textsc{compas} implements the interactions between the two stars of the binary systems during the different evolutionary phases (see \citealt{Compasetal2022} for more details). One such parameter is the natal kick velocity of the NS formed in the CCSN explosion of the heavier star in the binary system. We consider a NS that enters the RSG envelope at a late phase (Case B/C mass transfer), implying a relatively large orbital separation. A kick that is too large will unbind the NS-companion system.   

\begin{figure}
\begin{center}
\vspace*{0.15cm}
\hspace*{-0.45cm}
\includegraphics[width=0.53\textwidth]{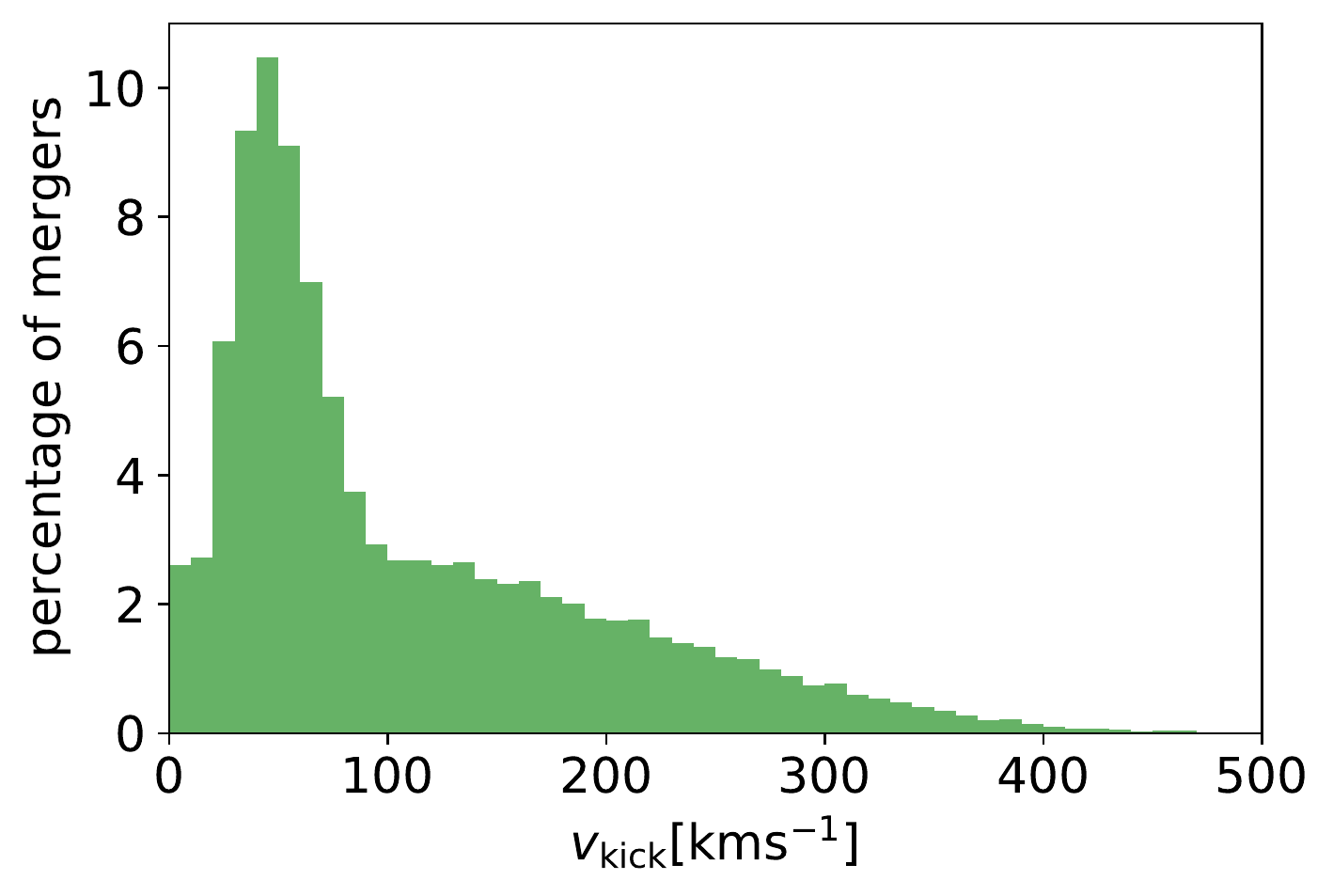}
%\vspace*{-5.5cm}
\caption{Natal-kick velocity distribution of the NS in system that results in NS-core mergers. The width of each bin is $10 \: \rm km \: s^{-1}$ }
\label{fig:v_kick_hist_1.00}
 \end{center}
 \end{figure}

Based on \cite{Hobbsetal2005} and \cite{PfahlRappaportPodsiadlowski2002} (see \citealt{Compasetal2022} for the implementation), \cite{Grichener2023} takes a bimodal natal-kick velocity distribution where the higher mode $\sigma_{\rm high}= 265 \km \s ^{-1}$ corresponds to regular, hydrogen-rich CCSNe and $\sigma_{\rm low}= 30 \km \s ^{-1}$ corresponds to ultra-stripped supernovae and electron-capture supernovae. In Fig. \ref{fig:v_kick_hist_1.00} we present the kick velocity distribution of the NSs in systems that result in NS-core mergers (whether ending in tidal disruption or in a CEE of the NS inside the core). We note a peak at $\simeq 50 \km \s^{-1}$, which is closer to the lower mode of the natal-kick distribution. Overall, the scenario we study here clearly favors slow natal kick velocities.
% FFFFFFFFFFFFFFFFFFFFFFFFFFFFFFFFFFFFFFFFFFFFFFFF
% FFFFFFFFFFFFFFFFFFFFFFFFFFFFFFFFFFFFFFFFFFFFFFFF

%=============================================
\section{A thermonuclear CEJSN scenario for SNR W49B}
\label{sec:W49B}
%=============================================

\begin{figure}
\begin{center}
\vspace*{-1cm}
\hspace*{-0.7cm}
\includegraphics[width=0.6\textwidth]{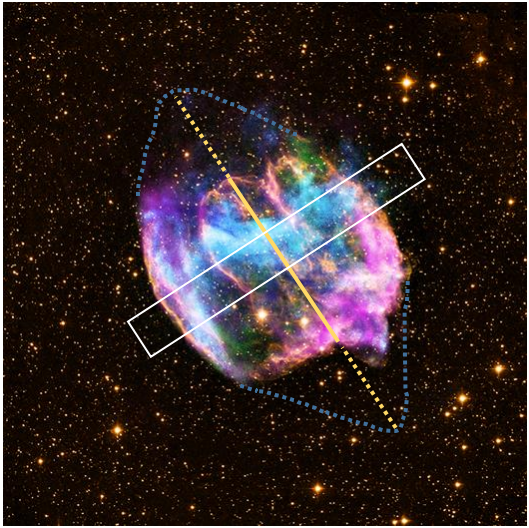}
\vspace*{-5.5cm}
\caption{Image of SNR W49B (from \citealt{Lopezetal2013b}; see also Chandra website  (http://chandra.harvard.edu/photo/2013/w49b/) with additional lines. The yellow line is the symmetry axis from \cite{BearSoker2017}, and the dotted cyan lines mark the two lobes that \cite{BearSoker2017} suggest to have existed in W49B (some segments of the lobes are still observed). 
We added the white rectangle to indicate the approximate equatorial plane of the NS-core system in the thermonuclear CEJSN scenario that we suggest for W49B. }
\label{fig:Lobes_SN}
 \end{center}
 \end{figure}

We propose that the SNR W49B is a result of a CEJSN triggered by tidal disruption of the RSG core. W49B is a barrel-shaped SNR (see Fig \ref{fig:Lobes_SN}) discovered over 60 years ago \citep{Westerhout1958}. However, the nature of its progenitor is still under debate. Some studies argue that the progenitor of W49B was a CCSN (e.g.,   \citealt{Lopezetal2011, Lopezetal2013b, Patnaudeetal2015}), while others claim that it was more likely a SNIa based on its metal abundances (e.g., \citealt{Hwangetal2000, ZhouVink2018, Siegeletal2020}). Nevertheless, there are still claims that W49B is a peculiar object that is hard to explain by any existing model (e.g., \citealt{Patnaudeetal2015, Siegeletal2020}).
% FFFFFFFFFFFFFFFFFFFFFFFFFFFFFFFFFFFFFFFFFFFFFFFF
% FFFFFFFFFFFFFFFFFFFFFFFFFFFFFFFFFFFFFFFFFFFFFFFF

We suggest that a CEJSN with a NS companion that tidally disrupts the CO core of an RSG star leading to thermonuclear ignition of the destroyed core material (henceforth we will refer to this scenario as `thermonuclear CEJSN') can account for various properties of W49B. The asymmetry of the metal distribution (e.g., \citealt{Siegeletal2021}) indicates that nuclear reactions occur before the core material formed an axially-symmetric accretion disk from the remains of the core around the NS. The shaping of the thermonuclear outburst in a non-axisymmetric disk might explain this highly asymmetrical metals distribution. The high energy of the Fe-K line (e.g., \citealt{Siegeletal2021}) requires a dense CSM that induces a strong reverse shock. As in planetary nebulae, we expect the pre-CEE binary interaction and the CEE to eject large amounts of mass and to form a dense CSM in all directions. The large-scale morphology of W49B can be explained by the interaction of the jets launched from the accretion disk with the dense ejecta from the destroyed RSG (\citealt{Akashietal2018}). The NS then collapses to a BH due to gas accretion from the core, and hence there is no evidence of a central compact object.

In table \ref{TAB:Table1} we list several properties of the SNR W49B inferred from observations (first column) and discuss how the previously proposed progenitors (second and third columns) and our proposed thermonuclear CEJSN scenario (forth column) might account for them. We also mark cases where there is no good explanation of how a scenario can account for a particular property.
% TTTTTTTTTTTTTTTTTTTTTTTTTTTTTTTTTTTTTTTT
% \usepackage{multirow}
\begin{table*}[]
%\hskip -3cm
%\small
\begin{tabular}{|p{4.0cm}|p{4.0cm}|p{4.0cm}|p{4.4cm}|}
% TTTTTTT CHANGE ONLY BELOW TTTTTTTTTTTT
\hline
\textbf{W49B Property} & \textbf{CCSN } & \textbf{SN Ia} & \textbf{Thermonuclear CEJSN}  \\ 
\hline
%{\color[HTML]{7030A0} Purple} & {\color[HTML]{CC0033} Red} & {\color[HTML]{385723} Green} & {\color[HTML]{0070C0} blue}  \\ 
Barrel-shaped$^{\rm [K07]}$ and H-shaped$^{\rm [L09,13]}$ morphologies& Jets launched by a NS $^{[\rm A18]}$ &\textcolor{red}{No explanation}$^{\rm [S19]}$& Jets launched by a NS that collapses to a BH $^{[\rm A18]}$ \\ 
 \hline
Concentrations of metals (Si/S/Ar/Ca/Fe) in an H-shaped$^{\rm [L13,Ch]}$ morphology & \textcolor{red}{Hard to explain if the symmetry axis is along the legs of `H'}$^{[\rm A18]}$& \textcolor{red}{No explanation}$^{\rm [S19]}$& Jets + Nuclear burning in a disk$^{\rm [TP23]}$ \\
 \hline 
High-luminosity and high-energy Fe-K line$^{\rm [P15,S21]}$
 & Dense CSM$^{\rm [Y14]}$& {Rare but possible with a high density CSM$^{\rm [S20]}$} & Like in the CCSN scenario. \\
 \hline
 Abundance of metals, e.g., $M_{\rm Fe} \simeq 1M_\odot$ $^{\rm [S20]}$  & \textcolor{red}{No good explanation}$^{\rm [S20]}$& Expected from an exploding massive white dwarf$^{\rm [S20]}$ & Burning of CO-rich gas in a disk around a NS$^{[\rm TP23ZB]}$ \\ \hline
No detection of a clear remnant$^{\rm [L13]}$ & BH remnant of the explosion & Expected & A BH remnant from accretion induced collapse of a NS$^{[\rm TP23]}$ \\
\hline
\end{tabular}
\caption{Comparison of three general scenarios for W49B. The last column is for the scenario proposed in this study. When referring to the H-shaped morphology we assume that the general (not accurate) symmetry axis is parallel to the legs of the `H', i.e., vertical through the center of `H' (\citealt{Akashietal2018}). 
References: A18: \citealt{Akashietal2018}; K07: \cite{Keohaneetal2007}; L09: \cite{Lopezetal2009}; L13: \cite{Lopezetal2013b};
P15: \cite{Patnaudeetal2015}; S19: \cite{Soker2019NewAR}; S20: \cite{Siegeletal2020}; S21: \cite{Siegeletal2021}; TP23: This paper; TP23ZB: This paper based on \cite{Zenatietal2020} and \cite{Bobricketal2022}; Y14: \cite{Yamaguchietal2014}; 
}
%TTTTTTTTTTTTTTTTTTTTTTTTTTTTTTTTTTTTTTTTTT
\label{TAB:Table1}
\end{table*}
\normalsize

The morphology of SNR W49B can serve as a benchmark for comparison between scenarios since there is a $90^{\circ}$ disagreement on the symmetry axis of this SNR. Here we take the symmetry axis of W49B's morphology to be as suggested by \cite{BearSoker2017} (and used also by \citealt{Akashietal2018} and \citealt{Siegeletal2020}) based on comparison of the structure of W49B to planetary nebulae. We mark the suggested symmetry axis (yellow line from \citealt{BearSoker2017}) and equatorial plane (the long axis of the rectangle) of the NS-core system on an image of W49B in Fig. \ref{fig:Lobes_SN}. The yellow line connects what \cite{BearSoker2017} assumed were the two polar lobes of SNR W49B that do not exist anymore. This axis is both the symmetry axis of the general X-ray barrel-shaped structure and of the H-shaped heavy metals (Si/S/Ar/CA/Fe; \citealt{Lopezetal2013b}) morphology. Namely, it is parallel to the legs of the `H' in the H-shaped distribution of the metals and vertical through the center of `H' (\citealt{Akashietal2018}). We assume that this is the symmetry axis of the two opposite jets that the NS/BH launches. This is in contradiction to the barrel shape axis \citep{Keohaneetal2007} and jets' axis (\citealt{Micelietal2008, Lopezetal2011, Lopezetal2013b, GonzalezCasanovaetal2014}) previously considered, which are perpendicular to this axis, i.e., along the high concentration of the iron and the rest of the metals (parallel to the rectangle we marked on Fig. \ref{fig:Lobes_SN}). Based on their hydrodynamical simulations and the latter choice for the symmetry axis, \cite{GonzalezCasanovaetal2014} argue that a jet-driven CCSN explosion can explain the distribution of metals, like silicon and iron, in SNR W49B.

 Further support to adopt the jets' axis parallel to the legs of the H-shape comes from the H-shaped structure of the X-ray emitting gas at the center of the galaxy M84 \citep{Bambicetal2023}.\footnote{see  https://phys.org/news/2023-05-m84-hot-huge-chandra-image.html for the identification of the H-shaped X-ray emitting gas in M84. } The radio emission clearly shows the direction of the jets to be parallel to the legs of the H-shaped gas, despite the fact the the jets' source is not at the center of the `H', as is the case in W49B.

The disruption of a white dwarf (WD) by a NS can trigger thermonuclear burning (outburst) in the accretion disk formed by the destroyed WD, leading to a transient event (e.g., \citealt{Fryeretal1998}; \citealt{Zenatietal2019}; see also \citealt{Yeetal2023} for thermonuclear burning together with jets in tidal disruption of a WD by an Intermediate-mass BH). \cite{Zenatietal2020} simulate the disruption of a WD by a NS and follow the nuclear burning in the disk around the NS. They simulate WD masses of $M_{\rm WD}=0.55M_\odot$, $0.63M_\odot$ and $0.8M_\odot$. The ejected iron group element (IGE) mass in the case of  $M_{\rm WD}=0.8M_\odot$ is $M_{\rm IGE}=0.045M_\odot$. This value is larger than for the other two masses by a factor of about 5. \cite{Bobricketal2022} simulate the disruption of a CO-rich WD with $M_{\rm WD}=0.9M_\odot$ and ONe-rich WDs with masses from $M_{\rm WD}=0.9M_\odot$ to $M_{\rm WD}=1.3M_\odot$. For the last case (of ONe $M_{\rm WD}=1.3M_\odot$) they find $M_{\rm IGE} \simeq 0.2 M_\odot$. 

From these we find the general trend in this mass range to very crudely be {$M_{\rm IGE} \approx 0.1 (M_{\rm WD}/M_\odot)^3 M_\odot $}. If this relation holds for NS-core interaction, it implies that a thermonuclear outburst from a disrupted core of $M_{\rm core} \simeq 2-3 M_\odot$ might yield the observed mass in W49B of $M_{\rm IGE} \simeq 1 M_\odot$. Our assumption requires confirmation by future 3D simulations of an RSG core disruption by a NS.

 Meanwhile, we can estimate the energetic in the system.
The nuclear burning of $M_{\rm Fe} \simeq 1M_\odot$ CO to iron accelerates the ejecta against the gravity of the NS to a kinetic energy of $E_{\rm Fe} \simeq 10^{51} \erg$. To shape this ejecta to an H-morphology the energy of the jets should be of the same order of magnitude. For jets propagating at the escape speed from a massive NS at $\simeq 0.5 c$, the mass in the jets that carry an energy of $10^{51} \erg$ is $M_{\rm jets} \simeq 3.6 \times 10^{-3} M_\odot$. The accretion disk that launches these jets has a mass about an order of magnitude larger, $M_{\rm disk} \simeq 0.03-0.1 M_\odot$.  Namely, only a small fraction, roughly $\simeq 0.01-0.1$, of the disrupted core is required to launch the jets after thermonuclear explosion.

We note that the NS tidally disrupts the core and starts to accrete mass from a thick torus that is not yet in a dynamical equilibrium. Namely, it has no axi-symmetric structure yet. This accretion disk might launch jets, but not those that shape the iron and other heavy metals that are synthesised later. The early jets runs ahead of the heavy metal ejecta. The NS can accrete mass at very high rates, up to $\simeq 1 M_\odot \s^{-1}$ as in CCSNe. After the NS accretes a large fraction of the core mass, i.e., $\simeq 0.5-2 M_\odot$, the disk reach high enough temperatures to ignite a thermonuclear outbursts that ejects most of the mass in the accretion disk, but not all of it. The leftover mass of $\sim 0.1 M_\odot$ is accreted by the massive NS or the newly born BH and launchers the final shaping jets.

%Mcore      M_{IGE}                        q    
%0.6          0.009 
%0.288             1.609     [ln(ratio)]     5.6         M_IGE=(Mcore)^q      
%0.8          0.045
%0.486        1.49                           3.1
%1.3          0.2 
%Predction
%2.5 -------> 1.5Mo
% ==========================================================
\section{Summary and discussion}
\label{sec:Summary}
% ==========================================================

In this work we proposed a new scenario of a thermonuclear CEJSN, where a transient event is powered and shaped by both jets and thermonuclear outbursts. In a thermonuclear CEJSN the core of a giant star gets tidally disrupted and forms an accretion disk around a NS (or a BH) that experiences thermonuclear burning. In addition, the accretion of core material through an accretion disk leads to energetic jets. 

We found the maximal core mass that leads to this scenario to be $2 M_{\rm \odot} \lesssim M_{\rm core} \lesssim 3.5M_{\rm \odot}$ for our chosen model parameters (equation \ref{eq:TidalConditionMass}). For heavier cores the NS enters a CEE inside the core. Based on this constraint on the core mass we inferred the initial RSG masses that lead to tidal disruption of the core by performing simulations with the stellar evolution code \textsc{mesa} (Fig. \ref{fig:MvsMcore}). Using then the data of NS-core mergers population synthesis from \cite{Grichener2023} we found the percentage of systems that result in tidal disruption of the core from all NS-core mergers (Fig. \ref{fig:mass_hist_percent_1.0} for core masses from \textsc{mesa}). From these we estimated that there is about one thermonuclear CEJSN with CO core per 1000 CCSNe, and about 0.2 CEJSN events where the NS enters a CEE inside a CO-rich core per 1000 CCSNe. The number of thermonuclear CEJSN deduced by taking the core masses from \textsc{compas} is slightly higher, by about $\approx 5-10\%$ .
 
In section \ref{sec:W49B} we proposed that the SNR W49B originated from a thermonuclear CEJSN event in which a NS tidally disrupted an RSG's CO core and an accretion disk was formed from the destroyed core material. The thermonuclear outburst took place before the accretion disk fully relaxed to an axisymmetrical disk, explaining the asymmetrical distribution of metal on the two sides of what we take as the symmetry axis of W49B (yellow line in Fig. \ref{fig:Lobes_SN}). The jets launched by the accretion disk can explain the general barrel-shaped structure and the H-shaped morphology of the metals. In our scenario the concentration of metals near the center is in the equatorial plane of the progenitor binary system (long axis of the rectangle in Fig. \ref{fig:Lobes_SN}). The event rate we found, of about one thermonuclear CEJSN with CO core per 1000 CEJSNe coincides with the fact that there is only one such SNR among the tens of observed CCSN remnants, and gives a prediction for future detection. We compare our proposed thermonuclear CEJSN scenario to the suggested CCSN and SN Ia scenarios of W49B in Table \ref{TAB:Table1}.  

Nucleosynthesis of r-process elements in the CEJSN r-process scenario  \citep{Papishetal2015} requires a merger of a NS with a CO core of an RSG \citep{GrichenerSoker2019a}. For the matter in the jets where the r-process occurs to be neutron rich enough and not mix with a large amount of iron this merger should occur during CEE of the NS inside the core, i.e., through scenario 1 (section \ref{subsec:TDEcriteria}).  Therefore, the rate of CEJSN that can produce r-process is about 0.2 in 1000 CCSNe. 
In \cite{GrichenerSoker2019a} we crudely estimate that a CEJSN event might synthesis $M_{\rm rp} \simeq 0.01-0.03 M_\odot$ of r-process elements. With the upper value of $ M_{\rm rp} \simeq 0.03 M_\odot$ and the rate we find here the CEJSN r-process scenario accounts for $\simeq 20\%$ of the r-process nucleosynthesis. However, our estimated values in \cite{GrichenerSoker2019a} were based on crude estimates and it is possible that we substantially underestimated that value. A larger r-process yield per CEJSN r-process event will lead to a higher fraction of r-process abundance from CEJSNe. Post processing of elaborated three dimensional hydrodynamical simulations can shed a light on the actual amount of r-process mass produced in a CEJSN event. 

%%%%%%%%%%%%%%%%%%%%%%%%%%%%%%%%%%%%%%%%%%%%%%%%
\section*{acknowledgments}
%%%%%%%%%%%%%%%%%%%%%%%%%%%%%%%%%%%%%%%%%%%%%%%%
We thank Alexey Bobrick for helpful discussions and Vikram Dwarkadas for helpful comments on the manuscript. We thank Dmitry Shishkin for providing the \textsc{mesa} stellar models. We thank an anonymous referee for comments and suggestions that helped in improving the manuscript. This research was supported by a grant from the Israel Science Foundation (769/20). A.G. acknowledges support from Irwin and Joan Jacobs Fellowship. Part of the simulations in this paper made use of the \textsc{compas} rapid binary population synthesis code (version 02.31.06), which is freely available at http://github.com/TeamCOMPAS/COMPAS.

\section*{Data availability}
The data underlying this article will be shared on reasonable request to the corresponding author.
%%%%%%%%%%%%%%%%%%%%%%%%%%%%%%%%%%%%%%%%%%%%%%%%

\end{document}